\documentclass[conference]{IEEEtran}
\IEEEoverridecommandlockouts
\usepackage{cite}
\usepackage{amsmath,amssymb,amsfonts}
\usepackage{algorithmic}
\usepackage{graphicx}
\usepackage{textcomp}
\usepackage{xcolor}
\def\BibTeX{{\rm B\kern-.05em{\sc i\kern-.025em b}\kern-.08em
    T\kern-.1667em\lower.7ex\hbox{E}\kern-.125emX}}
\begin{document}

\title{PReS: Power Peak Reduction by Real-time Scheduling for Urban Railway Transit}

\author{\IEEEauthorblockN{Zekun Yang, Yu Chen, Ning Zhou}
\IEEEauthorblockA{
\textit{Binghamton University, SUNY}\\
Binghamton, NY 13902, USA \\
\{zyang26, ychen, ningzhou\}@binghamton.edu}
\and
\IEEEauthorblockN{Shiqiong Tong}
\IEEEauthorblockA{\textit{Burns Engineering, Inc.} \\
Philadelphia, PA 19103, USA \\
STong@burns-group.com}
}

\maketitle

\begin{abstract}
Railway transportation is one of the most popular options for Urban Massive Transportation Systems (UMTS) because of many attractive features. A robust electric power supply is essential to enable normal operation. However, the power peaks appearing at the start time of the vehicles put heavy pressure on the power grid. Reduction of the power peak is a key issue in improving urban railway transit's power efficiency. Researchers have tried to address this problem by making a delicate timetable, but this method often failed to serve the purpose because of the punctuality problem. In this work, taking advantage of real-time estimation of the single train's power consumption, an online Power peak Reduction by real-time Scheduling (PReS) solution for trains' departure is proposed. Particularly, a Binary Integer Programming (BIP) model is introduced that is able to avoid power consumption peak caused by multiple trains departure simultaneously. The simulation result verified that the proposed real-time scheduling approach can effectively reduce the occurrences of power peak without bringing in additional train travel delay.

\end{abstract}


\section{Introduction}

While electrified railway transportation, considered as a convenient and reliable Urban Massive Transportation Systems (UMTS), will continue increasing rapidly in the world in the future, how to improve its energy efficiency is still a key issue \cite{scheepmaker2017review}. Besides energy consumption, the power peaks play a vital role in energy bills because of the infrastructure maintenance costs brought by instantaneous overload \cite{yang2016survey}. Generally, an extremely high traction power peak will occur when multiple trains start to accelerating simultaneously. Therefore, measures that avoid synchronized trains departure can effectively improve the energy efficiency and protect infrastructure in the power supply network.

In urban railway transit, coordinating acceleration of multiple trains by means of timetable (scheduling) optimization is a straightforward way. Following this philosophy, many researchers have proposed variant  approaches \cite{chen2005optimization,hu2013multi,min2010mathematical,huang2017joint}. However, the existing approaches do not necessarily operate well in practice. One of the challenges is that the timetable optimization approaches strongly depend on the punctuality of the trains, which is very difficult, if not impossible, to achieve, especially in the densely populated area during the busy hours. For example, the delay rate of the subway in New York City is as high as 65\%, which directly results in losing effect of the optimization. Furthermore, none of the existing approaches takes real power consumption data into account in the design of the timetable. Most of them are based on a simulative model, some even simply by assuming a constant power consumption, which of course introduces inaccuracy. 
  
In this paper, we propose a Power peak Reduction by real-time Scheduling (PReS) solution, an online scheduling strategy based on a real-time power consumption estimation to reduce the power peak incurred by multiple vehicles' simultaneous departure. It is non-trivial since it requires close cooperation of real time sensing, communication, computation and signaling systems in the urban railway system. Fortunately, Positive Train Control (PTC) system is implemented   \cite{zhao2015positive} to ensure the safety of railway traffic. 
The PTC will automatically stop a train before certain accidents occur, so it requires the railway system equipped with the functions of real-time track information sensing and data collection, GPS position information receiving on board, wireless communication among the control center, trains and wayside units, data computing and storage both on board and in control center. These functions, while built for PTC, can also enable the application of an online power consumption estimation and real-time data transmission to the control center. Several researchers have proposed approaches based on real time consumption estimation \cite{lesel2017real, rocha2018new}. Therefore, the proposed PReS scheme leverages the short-term power consumption estimation on board to dynamically reschedule trains' departure in control center to reduce power peak without large delay.    

The rest of this paper is organized as follows. Section \ref{sec:back} provides a brief review of the urban railway power system. Section \ref{sec:prop} presents the design of PReS including the high level framework, on-board power consumption estimation method and the scheduling optimization in control center. Section \ref{sec:exp} evaluates the performance of PReS through a case study. Conclusions are given in Section \ref{sec:con}.

\section{Urban Railway Power System}
\label{sec:back}
 
\subsection{Railway Power Supply Systems}

Railway system includes three different power supply systems: traction power system, railway signal system and auxiliary power supply system. The traction power system for electric trains’ operations is DC or single-phase AC networks. DC traction networks with voltages range between 600V and 3 KV are standard for urban mass transit of light rail systems. AC traction networks, normally 12 kV ~ 25 kV are utilized for high-speed and heavy haul rail systems. Three phase 60 Hz frequency utility power of 13.8 KV distribution level power or 69 KV, 115 KV, 230 KV transmission level power converted to 750 V DC or 12 KV, 25 Hz single phase to feed overhead catenary lines or third rails \cite{de2014railway}, which supply power for running electrical trains, are widely utilized traction power systems in the U.S. Railway signals normally have dedicated power supply system with backup power sources for reliability consideration and may have special voltage and frequency requirements. Auxiliary power supply is also needed for lighting, air conditioning, ventilation for railway tunnels, traction power substations and passenger stations.

For urban railway transit, DC traction power systems are standard due to the advantages of compact size and easy control with modern electronic technologies. The DC traction power system consists of traction power stations and power distribution conductors. 

\begin{itemize}
    \item Traction power substations (TPSS or TS), typically ratings in the range 1-10 MW, convert utility AC power to DC power through rectifiers for train operation. TPSSes are installed along railways with intervals about 1.6 miles (2.5km). The trains running on different sections of rails are powered by different TPSSes.
    \item Overhead catenaries or third rails are used as conductors to delivery power from TPSSes to trains. Running rails are used as electricity return path to TPSSes. Air gaps are kept on catenaries or rails as track section points to separate power supply sections for TPSSes. 
    \item Multiple passenger substations may be in one TPSS supply section. Trains’ acceleration or deceleration occurs frequently per schedule at passenger stations.
    \item Moving a train consumes power and the train’s accelerating procedure requests much higher power consumption than constant speed running.  Multiple trans may accelerate and run at one section at the same time. A TPSS is sized by the maximum power demands of trains within its power section. However, trains’ deceleration operation (the braking period of a vehicle) convert kinetic energy to electricity, which provide power to other trains or charge batteries of TPSSes.
\end{itemize}

\subsection{Analysis of Train Dynamics and Power Consumption}

\begin{figure}[t]
	\centering
	\includegraphics[width=0.42\textwidth]{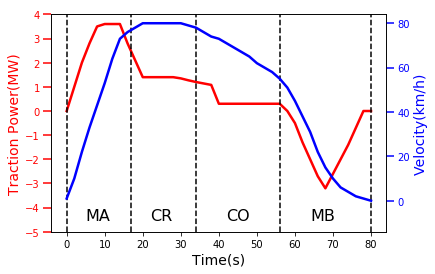}
		\vspace{-10pt}
	\caption{Four stages of train dynamics.}
	\label{fig:fig2}
	\vspace{-15 pt}
\end{figure}

 Typically, a duty circle of the train traveling from one station to another can be divided in four stages: maximum acceleration (MA), cruising (CR), coasting (CO), and maximum braking (MB) \cite{skogberg2015railway}. At first, the train departures from the station with the maximum acceleration as well as the maximum traction power. When the velocity increases to a certain value, the train shifts to the CR stage, which means operating with a constant speed. At this stage, the traction power stays at a low level to confront with the total resistance. At the third stage, the train starts to coast with zero traction power, and the velocity starts to decrease because of the resistance. Finally, the train begins stopping with the maximum braking force. If the train is equipped with regenerative braking, then the braking power will be converted into electricity to support other accelerating trains. Figure \ref{fig:fig2} illustrates the characteristics of the four stages. The required traction power comes to its maximum immediately after the train leaves the station.
 
 \begin{figure}[b]
 \vspace{-15 pt}
 	\centering
 	\includegraphics[width=0.42\textwidth]{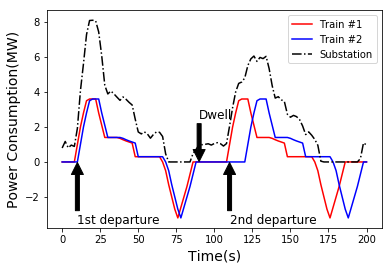}
 		\vspace{-10pt}
 	\caption{Example of multiple trains power consumption.}
 	\label{fig:fig3}
 	\vspace{-10 pt}
 \end{figure} 
 
 Figure \ref{fig:fig3} depicts the power curves of two trains operating in the same power range. The red and blue curves represent for train \#1 and \#2 respectively, and the black dotted line represents the total power consumption at substation level, in which a random value is added to total consumption for the auxiliary power consumption. When two trains leave the passenger station, they start to accelerate almost at the same time. Immediately an extremely high peak in the total power consumption appears after their departures. In contrast, when they stop and departure from the next station, the departure time is rescheduled. Train \#2 takes a longer dwell time than Train \#1, and as a result, the total power consumption does not experience the same high spike. This illustration shows the intuition of the dynamic rescheduling, avoiding multiple trains departure simultaneously effectively reduces the power consumption peak.

\section{Real-time Scheduling for Power Peak Reduction} 
\label{sec:prop}

\subsection{PReS System Level View}

\begin{figure}[t]
	\centering
	\includegraphics[width=0.48\textwidth]{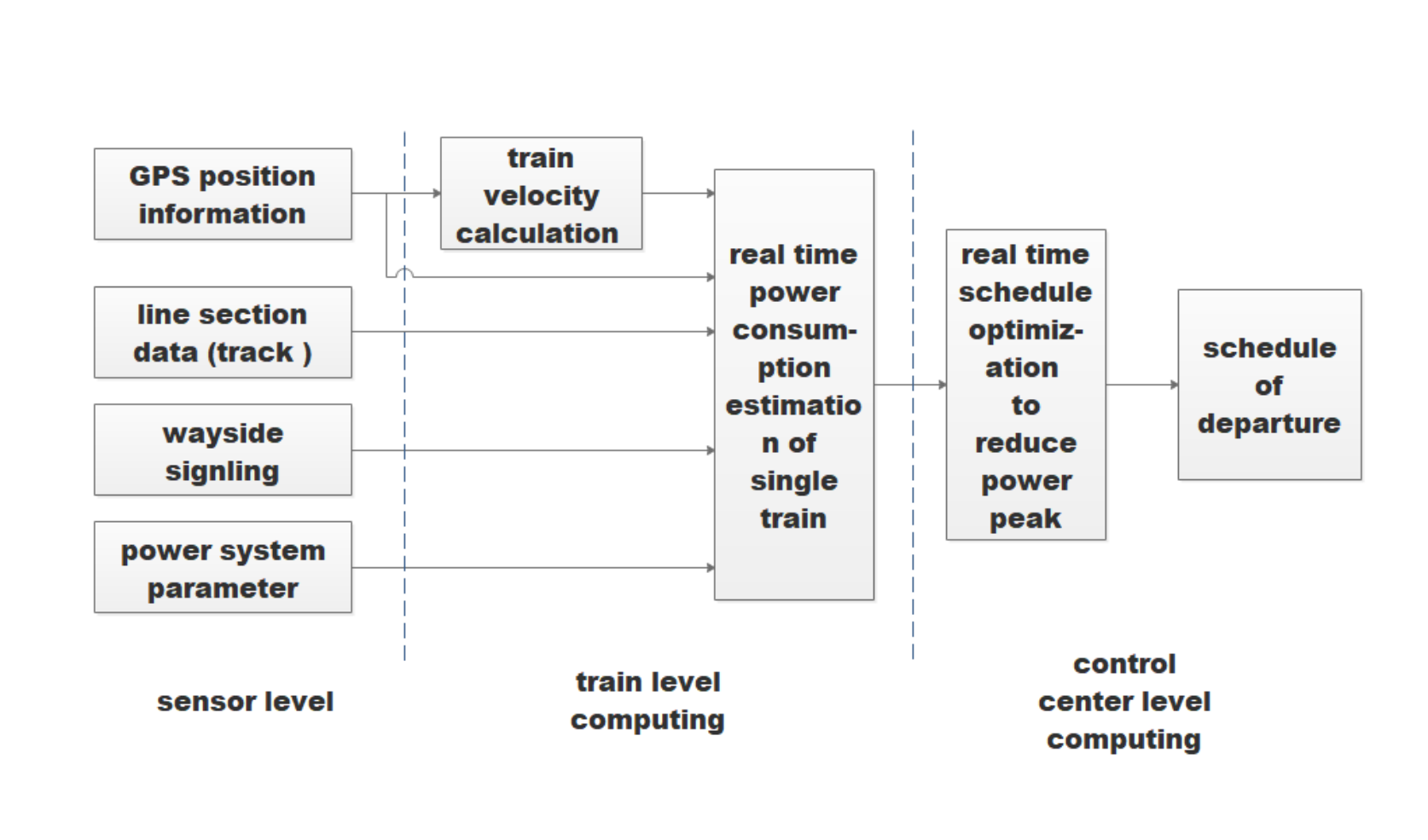}
	\caption{PReS approach framework.}
	\label{fig:frame}
	\vspace{-15 pt}
\end{figure}

Figure \ref{fig:frame} shows the proposed PReS system following a hierarchical architecture \cite{chen2016smart}, \cite{nikouei2018smart}. At the sensor level, multiple data are collected by the sensors to obtain the real-time status of the train, including the GPS position, the line section, speed, power system status, etc. The data are processed on-site at the train for the power consumption estimation, such that there is not any unnecessary data traffic on the communication network, also the communication delays are avoided. Then the estimation result is transmitted to the central office, where the departure time scheduling is conducted on top of the estimations from vehicles in the close geographic area.  


\subsection{On-board Power Consumption Estimation}
The motion of a train is described by the Newton's equation:
\begin{equation}
m\cdot a=\alpha f(v)-\beta b(v)-F_m(v)-F_g(x)-F_c(x)
\end{equation}
\begin{equation}
\hspace{-6.5cm}
\frac{dx}{dt}=v
\end{equation}
\begin{equation}
\hspace{-6.5cm}
\frac{dv}{dt}=a
\end{equation}
where
\begin{quote}
\begin{itemize}
	\setlength{\leftmargin}{2cm}     
	\item[$m$] the static mass of the train
	\item[$a$] the acceleration of the train
	\item[$v$] the velocity of the train
	\item[$x$] the position of the train
	\item[$f(v)$] the maximum traction force
	\item[$b(v)$] the maximum braking force
	\item[$\alpha$] the factor of the traction force, $0\le \alpha \le 1$
	\item[$\beta$] the factor of the traction force, $0\le \beta \le 1$
	\item[$F_m(v)$] the train resistance to motion
	\item[$F_g(x)$] the train resistance due to the track gradient
	\item[$F_c(x)$] the train resistance due to the track curve

\end{itemize}
\end{quote}

When the train is operating in the MA or CR stage, the tractive effort are pushing the train forward, therefore the factor of the traction force $\alpha > 0$ and the factor of the braking force $\beta = 0$. During the CO stage, both traction and braking force are zero, i.e., $\alpha =0, \beta = 0$. When the train is arriving a station, only braking force works to help the train stop, therefore, $\alpha =0, \beta > 0$. In practice, the tractive effort does not necessarily keep the maximum value, because of some operational considerations, such as the passengers' comfort. 

\begin{figure}[tb]
	\centering
	\includegraphics[width=0.41\textwidth]{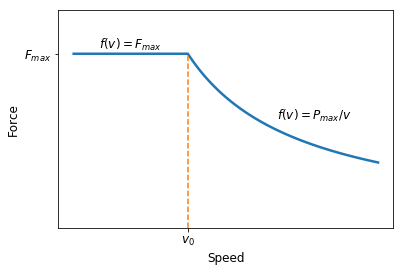}
	\caption{Force-vs-speed curve.}
	\label{fig:fig5}
	\vspace{-15 pt}
\end{figure}

Figure \ref{fig:fig5} is a force-vs-speed curve, which shows that the maximum tractive effort produced by the locomotive is velocity-dependent. When the velocity is below a certain level, the traction is constrained by the maximum torque, so the maximum traction keeps a constant. When the velocity gets higher, the traction is limited by the maximum power $F(v)\times v$, so the maximum traction is inversely proportional to the velocity. With the identified curve trend, each rolling stock has a particular force-vs-speed diagram, which is provided by the manufacturer. The maximum braking force also depends on the velocity, similar to the force-vs-speed relationship. 

For the train resistance to motion $F_m(v)$, Davis and the American Railway Engineering Association proposed the Davis equation \cite{lukaszewicz2007running}, which has been applied by many train simulators and analysis models. According to the Davis equation, the resistance of motion is defined as:
\begin{equation}
F_m(v)=6.4m+130n+0.14mv+\gamma[0.046+0.065(N-1)]Av^2
\end{equation} 
where
\begin{quote}
\begin{itemize}
\item[$m$] the static mass of the train
\item[$n$] the number of axles
\item[$v$] the velocity of the train ($km/h$)
\item[$N$] the number of cars
\item[$A$] the front section area of the train
\item[$\gamma$] a coefficient, 1 if the train is outside, and 2 or 3 if in tunnels 
\end{itemize}
\end{quote}

The gradient force $F_g$ is calculated as $F_g(x)=mg\sin\theta$, where $\theta$ is the track gradient angle. The train resistance due to the curve $F_c$ is calculated as $F_c(x)=\frac{ke}{r} 10^{-3}mg$, where $ke$ is a coefficient of the track gauge, and $r$ is the radius of the curve. If the instantaneous position and velocity of the train are given, it is easy to get the traction power $\alpha vf(v)$ or the braking power $\beta vb(v)$ from equations above. 


\subsection{Real-Time Train Departure Scheduling}

The problem of real-time scheduling can be formatted as a train departure decision-making problem at each time instant. At the $i$th time slot, there are a group of trains $K_i$ dwelling at passenger stations in a particular substation range. They are ready to departure and waiting for the authorization from the control center. The algorithm should give a solution of $K_i=\{k_{ij}| j\in M\}$, where $k_{ij}=1$ means vehicle $j$ is allowed for departure, while $k_{ij}=0$ means vehicle $j$ still needs to dwell at the station and wait. 

\textbf{Inputs:}
\begin{quote}
	\begin{itemize}
		\setlength{\itemsep}{3pt}
		\item[$t_i$] current time
		\item[$N_i$] set of vehicles operating in the power range in the time instance $t_i$, $N_i=(1,2,\dots,n)$
		\item[$h_{ij}^1$] headway of vehicle $j$ with the previous vehicle, $j\in N$
		\item[$h_{ij}^2$] headway of vehicle $j$ with the following vehicle
		\item[$M_i$] set of vehicles waiting for departure
		\item[$P_{ij}$] power consumption of vehicle $j$ at the time instance $t_i$. $P_{ij}=0$ when $j\in M_i$, positive when vehicles need traction force, while negative when vehicles are regenerating power.  
	\end{itemize}
\end{quote}

\textbf{Parameters:}
\begin{quote}
	\begin{itemize}
		\item[$T_j$] pre-designed schedule of vehicle $j$
		\item[$v_j$] maximum velocity of vehicle $j$
		\item[$p_j$] estimation of instantaneous power of vehicle $j$'s departure
		\item[$h_{min}$] constraint of minimum headway between vehicles
		\item[$\Delta t$] time interval
		\item[$P_{th}$] the threshold of the total traction power
	\end{itemize}
	\end{quote}
	
\textbf{Outputs:}

 $K_i=\{k_{ij}| j\in M\}$.

\textbf{Constraints:}
\begin{enumerate}
	\item Vehicles waiting for departure belong to vehicles operating in the power area:
	\begin{equation}
		M_j \subseteq N_j
	\end{equation}
	
	\item Vehicles shall not collide with any other vehicles currently at the station:
	
	\begin{equation}
	h_{ij}^1-k_{ij}\cdot \Delta t \cdot v_j\ge h_{min}
	\end{equation}
		
	\item Vehicles shall not collide with departed vehicles:
	
	\begin{equation}
	h_{ij}^2-(1-k_{ij})\cdot \Delta t \cdot v_j\ge h_{min}
	\end{equation}
	
	\item The total traction power cannot surpass a given threshold. Since the regenerative power cannot flow back to the transmission network through the TS, without additional energy storage equipment, it can be utilized by the nearby vehicles for acceleration. If the regenerative braking energy is not used immediately, the regenerative brake will be converted into a resistance brake and air brake, and the braking energy is wasted in vain. The total traction power of all vehicles operating in the power range can be calculated as Eq. \ref{eq:ttlpw}:
	
\end{enumerate}

	\begin{equation}	
	P_{total,i}=
	\left\{
	\begin{array}{c}
	\renewcommand{\arraystretch}{}
	\sum\limits_{j=1}^{n}P_{ij}+\sum\limits_{j=1}^{m}p_j\cdot k_{ij},  \mathrm{if}-\sum\limits_{j=1}^{n, P_{ij}\le 0}P_{ij}\le \sum\limits_{j=1}^{m}p_j\\
	\\
	\sum\limits_{j=1}^{n, P_{ij}\ge 0},\qquad \qquad \mathrm{if}-\sum\limits_{j=1}^{n, P_{ij}\le 0}P_{ij}> \sum\limits_{j=1}^{m}p_j\\

	\end{array}
	\right.
	\label{eq:ttlpw}
	\end{equation}

To reduce the peak power consumption means to keep the total power consumption $P_{total,i}$ below the threshold. However, if there exists one or more vehicles owning an unusual high individual departure power $p_j$, we cannot stop it from departure simply because it surpasses the power threshold. In another word, the limitation of the total power consumption should be a "soft" constraint. Thus, two adjustment factors $d^-$ and $d^+$ are introduced to the constraint condition.

\begin{equation}	
\left\{
\begin{array}{l}
	P_{total,i} + d^- + d^+=P_{th}\\
	\\
	d^-\ge 0, d^+\ge 0
\end{array}
\right.
\end{equation} 

In summary, the constraint conditions are defined as follow: 	

\begin{equation}	
\left\{
\begin{array}{l}
M_j \subseteq N_j\\
\\
h_{ij}^1-k_{ij}\cdot \Delta t \cdot v_j\ge h_{min}\\
\\
h_{ij}^2-(1-k_{ij})\cdot \Delta t \cdot v_j\ge h_{min}\\
\\
P_{total,i} + d^- - d^+=P_{th}\\
\\
d^-\ge 0, d^+\ge 0
\end{array}
\right.
\end{equation}


The objective of the optimization is to minimize the total vehicle delay with the constraints of power consumption and meet the safety consideration. The total vehicle delay at the time instant is:

\begin{equation}
	\mathrm{Total\quad delay} = \sum_{j=1}^{M}[t_i-T_j+(1-k_{ij})\Delta t]
\end{equation}

To simplify the question, the objective is dimidiate: on the one hand, it is desired to maximize the number of vehicles departure at a time, i.e.,$\sum_{j=1}^{M} k_{ij}$; on the other hand, it is also expected to maximize the utilization of energy in the possible range of the power consumption. The objective is defined as:
   
\begin{equation}
	\mathrm{\textbf{min}}\qquad w_1 d^+ -w_2 \sum_{j=1}^{M} k_{ij}
\end{equation}

\noindent where $w_1$ and $w_2$ are weights of the dimidiate parts respectively. 
While $w_1$ emphasizes the importance of the total power consumption, $w_2$ prevents too much sacrifice of the train travel time. In practice, two correction factors are added, $\gamma_1$ and $\gamma_2$, to the objective function. $\gamma_1$ increases the usage of regenerative energy, such that $\gamma_1 > 0$ when $P_{total,i} < 0$ and $\gamma_1 = 0$ when $P_{total,i} \ge 0$. Meanwhile, $\gamma_2$ encourages newly available trains to departure at the current time point. According to the real-time optimization method, trains tend to postpone departure in order to avoid power peak appear at the current time point. But this may result in power peak in the future and unnecessary delays. Therefore, another correction parameter $\gamma_2$ is introduced that is directly proportional to the number of trains that are newly available to departure. Finally, the objective function is changed to:    

\begin{equation}
	\mathrm{\textbf{min}}\qquad w_1 d^+ -(w_2 + \gamma_1 + \gamma_2) \sum_{j=1}^{M} k_{ij}
\end{equation}

The possible value of $k_{ij}$ is 1 or 0, so this optimization problem is solved as a Binary Integer Programming (BIP) question. Since the single train power consumption has already been calculated on board using real-time measurement data, the optimization is simplified with linearity. Therefore, compared with other existing solutions using time-consuming optimization algorithms like Genetic Algorithm (GA), our problem can be solved by a basic algorithm like Balas Additive Algorithm, and feasible to real-time application. 

\section{Experimental Study}
\label{sec:exp}

\subsection{Experimental Setup}
To validate the effectiveness of the proposed PReS model and the numerical algorithm, a simulation experiment is conducted based on a simplified scenario of urban railway transit system. The system is formulated as a discrete event simulation and realized by Python SimPy package, and the BIP optimization problem is solved by Python PuLP package.  

\begin{figure}[b]
   \vspace{-15pt}
	\centering
	\includegraphics[width=0.4\textwidth]{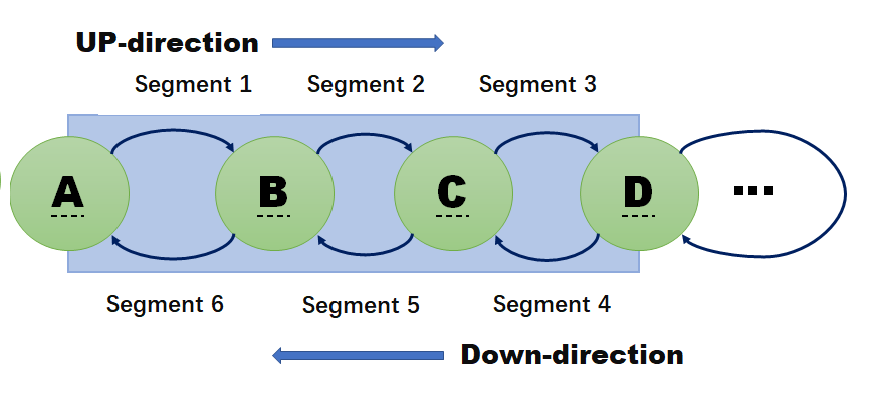}
		\vspace{-10pt}
	\caption{Simulation scenario of train route.}
	\label{fig:Route}
	\vspace{-10pt}
\end{figure}

In order to study the power supply of a single station, a bi-direction metro corridor with four railway stations is considered. Figure \ref{fig:Route} shows that there are six segments among four stations, including both up-direction and down-direction, which constitute the coverage area of a single power supply substation. In the simulation, trains start from station A, travel along the up-direction path A $\to$ B $\to$ C $\to$ D with the average segment travel time [180s, 120s, 150s], travel out of the current substation coverage area with an average period 1080 seconds, then go back to the down-direction path D $\to$ C $\to$ B $\to$ A the average segment travel time [150s, 120s, 180s], and finally finish the journey by arriving at station A. The railway system has a fixed headway of six minutes (360s), which means every six minutes there is a train leaving the departure station A. The minimum distance between two adjacent trains is three minutes (180s). The passenger time at each station is one minute, such that train dwell time at each stations should be at least 60 s.   

As a discrete event model, the system time interval is ten seconds. In every 10 s, a single train traction power usage is calculated and the operator center makes a decision of which train is allowed to departure. For simplicity, the calculation of the total power consumption of a substation only considers the traction power usage, since the auxiliary power consumption does not change with the real-time scheduling. And the single train traction power for every segment depends on a uniform profile of [10, 8, 8, 2, 2, 3, 2, 2, 2, 0, -6, -8, -4] $\times 10^3$ kW, and scales according to the segment travel time.   

Two scenarios are experimentally studied. One with a fixed timetable and the other one with real-time scheduling, with a total simulation time 20000 seconds. For the scenario with real-time scheduling, the parameter selection of the BIP optimization is as the follows: two weights $\omega_1 = 3$, $\omega_2 = 5$, two corrections $\gamma_1 = 20$, $\gamma_2 = 1 \times $ number of newly available trains, and the power consumption threshold is $125 \times 10^3$ kW.

\subsection{Experimental Results}

\begin{figure}[b]
    \vspace{-15pt}
	\centering
	\includegraphics[width=0.4\textwidth]{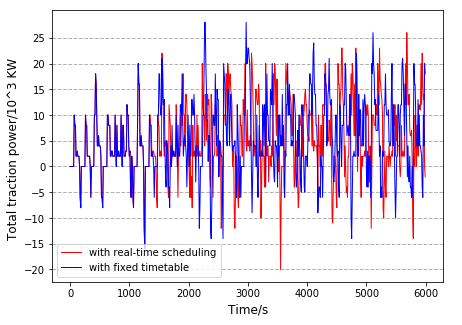}
		\vspace{-10pt}
	\caption{System power consumption profile in the first 6000s.}
	\label{fig:PowerRecord}
	\vspace{-10pt}
\end{figure}

Figure \ref{fig:PowerRecord} shows the system power consumption profile in the first 6000 s of the simulation. In the beginning 1500 s, when trains operate in up-direction and the total power consumption is low, the PReS system does not make any real-time rescheduling to change the pre-set timetable. Hence two scenarios present the same power consumption profile. After 1500 s, the total power consumption increases, the PReS function is triggered and the real-time rescheduling effectively reduces the frequency of power peak occurrence. Given the power consumption threshold of $25 \times 10^3$ kW, the scenario with fixed timetable generates power peak above the threshold three times, while the scenario with real-time scheduling only has one power peak above the threshold.  

\begin{figure}[t] 
  \centering
  \includegraphics[width=0.5\textwidth]{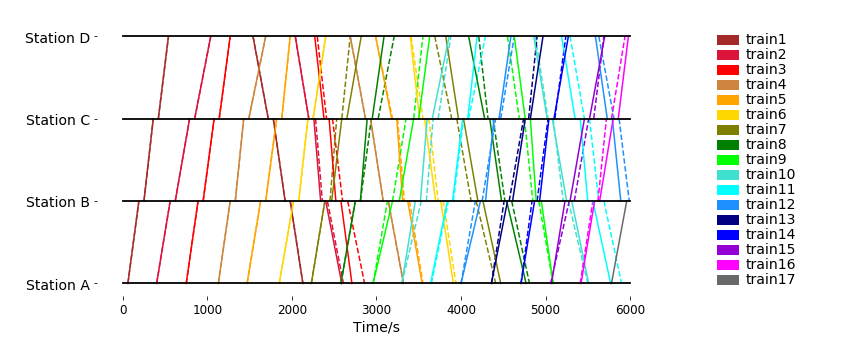}
  \caption{Train scheduling in first 6000 s.}
  \label{fig:timetable}
  \vspace{-15pt}
\end{figure}

The dynamic rescheduling may introduce additional delays, thus the travel time is an important factor. Figure \ref{fig:timetable} shows the train schedule in the beginning 6000 s. The solid line represents the travel schedule of trains following the pre-fixed timetable, while the dotted line represents the travel schedule of trains with real-time scheduling. Similar to the power consumption profile, in the beginning 1500 s, trains in both scenarios operate identically. After the 1500 s, trains start to operate with differences. Sometimes trains travel faster if they follow the fixed timetable, such as train 3; sometimes trains travel faster if they follow the dynamic scheduling given by the PReS, such as train 7. Considering the results shown in Fig. \ref{fig:PowerRecord} along with Fig. \ref{fig:timetable}, the proposed PReS real-time scheduling method reduces the power peak occurrence time without introducing an obviously extra delay.   

\begin{figure}[b]
   \vspace{-15pt}
	\centering
	\includegraphics[width=0.4\textwidth]{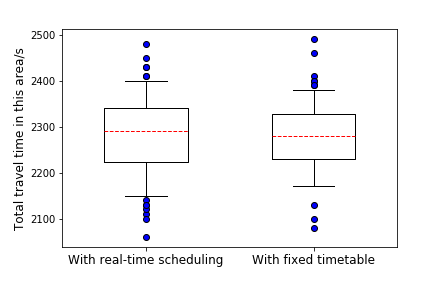}
		  \vspace{-10pt}
	\caption{Box plot of single train travel time.}
	\label{fig:TravelTime}

\end{figure}

Figure \ref{fig:TravelTime} reports the travel time of totally 134 trains. 
The two scenarios have similar average and deviation. Table 1 summarizes the comparison of the two scenarios. When the trains operate following the PReS scheme with real-time scheduling, the power peak that surpasses the power consumption threshold is reduced from 28 times to 8 times (reduced 71.4\%). Meanwhile, because of the correction of regenerative power, the PReS scheme also shows a minor advantage in utilizing the regenerative power. Because of the higher power usage efficiency, the PReS system also consumes less power in average per second. As a tradeoff, while the PReS scheme brings several advantages, it may introduce some extra delays. However, the introduced delay is tolerable as shown by the simulation result. In terms of the average total travel time of 40 minutes (2400 s) for 134 trains, the PReS system introduced about nine seconds of extra delay comparing to the operation with a pre-fixed timetable, which is trivial (0.38\%).

\section{Conclusions}
\label{sec:con}

In this paper, we proposed a PReS scheme consisting of a mathematical train traction power consumption model and a BIP optimization algorithm, which helps to reduce the occurrence of power peaks at substations. With an online scheduling based on the real-time train power estimation, the operator is allowed to adjust the timetable corresponding to the current situation. The experimental results validated the proposed PReS scheme by showing an effective reduction of 71.4\% of the power peaks that surpass the reference threshold with a trivial overhead of 0.38\% of extra delay time. In addition, the PReS scheme also helps to improve the efficiency of the entire railway system, i.e. better usage of regenerate power and lower average power consumption.

\begin{figure}[t]
   \vspace{-10pt}
	\centering
	\includegraphics[width=0.45\textwidth]{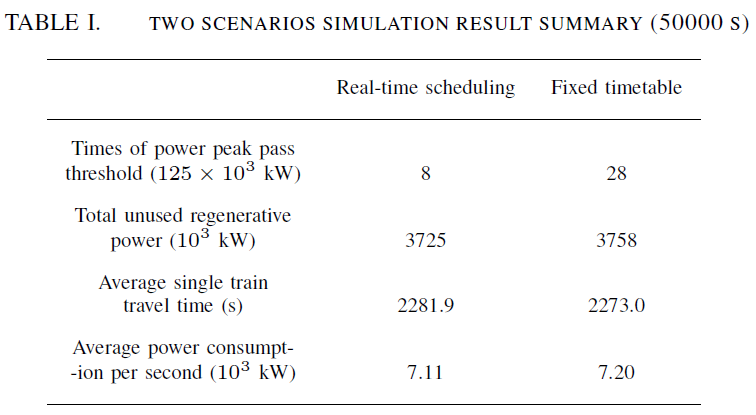}
	\label{fig:table1}
	  \vspace{-15pt}
\end{figure}

\bibliography{ref}
\bibliographystyle{IEEEtranS}

\end{document}